\documentclass[pre,twocolumn,groupedaddress,showpacs,floatfix,notitlepage,color]{revtex4-1}
\usepackage{graphicx}
\usepackage{epsfig}
\usepackage{amsfonts}
\usepackage{color}
\usepackage{ulem}
\usepackage{amsmath}
\usepackage{mathrsfs}
\usepackage{subfigure}
\usepackage[autostyle]{csquotes}
\usepackage[utf8]{inputenc}
\usepackage[T1]{fontenc}
\usepackage{lmodern}

\begin{document}

\title{Stokes Waves Revisited: Exact Solutions in the Asymptotic Limit}

\author{Megan Davies}
\affiliation{Formerly: Aston University, Chemical Engineering and Applied Chemistry, Aston Triangle, Birmingham B4 7ET, UK}
\email{davies@megan4.orangehome.co.uk}
\author{Amit K Chattopadhyay}
\affiliation{                    
Aston University, Mathematics, Aston Triangle, Birmingham B4 7ET, UK}
\email{a.k.chattopadhyay@aston.ac.uk}

\begin{abstract}
Stokes perturbative solution of the nonlinear (boundary value dependent) surface gravity wave problem is known to provide results of reasonable accuracy to engineers in estimating the phase speed and amplitudes of such nonlinear waves. The weakling in this structure though is the presence of aperiodic \enquote{secular variation} in the solution that does not agree with the known periodic propagation of surface waves. This has historically necessitated increasingly higher ordered (perturbative) approximations in the representation of the velocity profile. The present article ameliorates this long standing theoretical insufficiency by invoking a compact exact $n$-ordered solution in the asymptotic infinite depth limit, primarily based on a representation structured around the third ordered perturbative solution, that leads to a seamless extension to higher order (e.g. fifth order) forms existing in the literature. The result from this study is expected to improve phenomenological engineering estimates, now that any desired higher ordered expansion may be compacted within the same representation, but without any aperiodicity in the spectral pattern of the wave guides. 
\end{abstract}
\date{\today}


\maketitle

Deep water surface gravity waves are conspicuous in their typical crest-trough nonlinear patterns which show periodicity in their spectral pattern \cite{grant1973,svendsen2006}. The generic issue confronting theoretical solutions of associated models is that of the boundary condition dependence of related nonlinear, but often periodic, waveguides. This difficult problem was tackled by Stokes in what then became a high point of the success of mathematical analysis in explaining fluid wave propagation. Using a Taylor series expansion around the mean surface height profile, the later day equivalent of a linear stability analysis \cite{jordansmith2007}, Stokes was able to derive a perturbative solution of the wave velocity, using {\it wave steepness} as a measurable perturbation parameter. The result in turn led to a formal quantitative explanation of phase speeds and amplitude spectra observed in coastal waves, including advected tidal waves generated by the motion of ships. 

Stokes analysis had some restrictions though. While being reasonably accurate in deep water surroundings, at shallow water, characterized by a large wavelength  ($\lambda$) to mean depth ($h$) ratio ($r=\frac{\lambda}{h}>>1$), the perturbative Stokes solution breaks down. This is not very difficult to perceive either. A large value of the wavelength:depth ratio $r$ effectively implies a large enough value for the wave steepness at which point the very nature of a perturbative analysis becomes at stake, eventually breaking down. Later modified theories using a Boussinesq approximation (instead of the initial Poincare-Lindstedt method used by Stokes \cite{poincare1957}) improved the quantitative match but the solitonic solutions \cite{higgins1974} still remained limited to the deep to intermediate water depths. Two major works on finite depth Stokes waves were firstly the third order \cite{liu1989,dingemans1997} and later the fifth order theories \cite{de1955} that calculated the phase speed (celerity) up to fifth order of accuracy.

A series of next generation breakthroughs in this lineage resulted in extending De's fifth-ordered perturbative solution of the Stokes form by Fenton \cite{fenton1985} and comparable {\it cnoidal wave} theories \cite{hedges1995}. All these analyses relied on close association of statistical modelling to phenomenological studies that were gratifying to engineers who depended on numbers to hardgrind their estimates but from a theoretical perspective, there were two unfounded issues that demanded explanation. Firstly, while convergence of the Stokes expansion could be proved in the infinite ranged expansion \cite{debnath2005}, a finite ordered small amplitude theory had a closure issue, leading to a lack of convergence. In other words, a compact representation of the Stokes' wave formulation for the infinite depth situation is still lacking. Secondly, even in the deep water limit, Stokes waves were shown to be unstable \cite{zakharov2009}. This instability is known in the literature as a Benjamin-Feir instability and arises due to side-band modulations of the propagating surface waves. The technical issue with such an instability is the fact that the instability arises from a nonlinearity in the structure leading to a model that can be mapped onto a nonlinear Schr{\"o}dinger equation, which then can only be solved approximately analytically, or else numerically only. Also this model lacks a generic periodic solution for most boundary conditions. The lineage of approximate perturbative solutions based on Stokes original model led to a series of relevant computer modelling works as well \cite{craig2002}, a consummate summary of which is available in the book by Mader \cite{mader2009}. In this work, we will address the asymptotic infinite depth Stokes' theory to obtain a closed form compact solution of the velocity, acceleration and kinetic energy representations for any perturbative order.

The problem of Stokes surface waves, leading to Stokes turbulence at low Reynold's number, is a classic boundary layer problem. When a periodically ramped flow hits a solid wall, or else if an oscillating plate moves relatively in a viscous fluid at rest, the fluid boundary layer close to the solid wall assumes a nonlinear profile driven by non-inertial forces. In a seminal work, Stokes showed that such oscillatory flows give rise to boundary-layer eddies close to the boundary wall, away from which they decay exponentially to a stable inertial regime \cite{landau1987}.

In this article, we will provide a simple yet compact, and importantly, converging solution of Stokes gravity wave model for free surface boundary conditions. Our generic solution can also be extended to the oscillating pressure gradient regime, as also in most other waveforms that admit of progressive wave solution. Our formulation starts with the progressive wave hypothesis for the propagation vector $z$ that defines the free surface elevation in the (x,y) plane: $\eta(x,t) = \eta(x-ct)$ and ${{\bf u}}(x,z,t)={{\bf u}}(x-ct,z)$. Defining $\theta(x,t)=kx-\omega t=k(x-ct)$ as the spatially varying wave phase, where phase velocity $c=\frac{\omega}{k}$, the free surface elevation $\eta(x,t)$ and the velocity potential $\phi(x,z,t)$ can be represented through Fourier series sums:

\begin{subequations}
\begin{equation}
\eta(\theta,x,t) = \displaystyle \sum_{n=1}^{\infty} A_n\:\cos(n \theta)
\label{eta-fourier}
\end{equation}
\begin{equation}
\phi(\theta,x,z,t) = \beta x - \gamma t +\displaystyle \sum_{n=1}^{\infty} B_n \lbrack{\cosh(nk(z+h))}\rbrack\:\sin(n\theta),
\label{phi-fourier}
\end{equation}
\end{subequations}

where $h(x,y)$ relates to the normal component of the flow velocity defined through the relation $\frac{\partial \phi}{\partial z}=0$ at $z=-h$. As to the form of the constants $A_n$'s, the third order Stokes solution \cite{liu1989,dingemans1997} is indicative:

\begin{eqnarray}
\eta(\theta, x, t) &=& a\bigg[ \cos(\theta)+\frac{1}{2}(ka)\cos(2\theta)+\frac{3}{8}{(ka)}^2\cos(3\theta)\bigg]\nonumber \\
&+& O({(ka)}^4),
\label{third-order}
\end{eqnarray}

in which $a$ is the first-order wave amplitude and $\theta$ is the wave phase. What we do now is to hypothesise an $n$-order generalization based on this third order formulation and later show that this conforms to the fifth order solution (Figures 1-3). The alluded $n$-ordered representation is proposed as follows:

\begin{equation}
\eta(\theta,x,t)=\frac{\zeta}{k} \cos(\theta) + \displaystyle \sum_{n=2}^{\infty} \zeta^{n} \left(\frac{n}{2^{n}}\right) \:\cos(n\theta),
\label{n-order}
\end{equation}

where $\zeta^2=ka$, $\zeta^3={(ka)}^2$, etc., {\it i. e.} $\zeta=ka$. The formulation in equation (\ref{n-order}) has the advantage that it is perturbatively accurate up to any higher order, for example to the third \cite{dingemans1997} or to the fifth order \cite{fenton1985} expansions. It must be noted that this $n$-ordered representation as suggested in equation (\ref{n-order}) is not an {\it ab initio} deduction, rather this is based on a correct comprehension of the underlying symmetry in the third order perturbative solution, that eventually continues to higher orders (we have checked up to the seventh ordered form) with reasonable levels of accuracy.

Comparing the two expressions in equations (\ref{third-order}) and (\ref{n-order}), the latter up to $n=5$, we can see that in the Stokes equation formulation, the third and fourth terms in equation (\ref{third-order}) might be combined, the focal term here being the $\cos(3\theta)$ harmonics in equation (\ref{third-order}). Perturbatively, the amplitudes from the higher ordered terms e.g. fourth, fifth and sixth terms will have negligible effect on an observed wave height. Equation (\ref{n-order}) can be exactly solved to obtain a converging solution for all higher ordered wave forms to a high level of accuracy, resulting in the following solution

\begin{eqnarray}
\eta_{n\to \infty} &=& \frac{\zeta e^{2ix} [(\zeta^4 + 24\zeta^2 + 32) \cos(\theta)}{2{(\zeta-2e^{ix})}^2{(\zeta e^{ix}-2)}^2}\nonumber \\
 &-& \frac{4\zeta (8+\zeta^2 + (4+\zeta^2)\cos(2\theta)-\zeta\cos(3\theta))]}{2{(\zeta-2e^{ix})}^2{(\zeta e^{ix}-2)}^2}
\label{vel-inf}
\end{eqnarray}

The importance of the third harmonic ($\cos(3\theta)$) term relates to the origin of this presentation; while the form above does not uniquely prove the level of convergence to higher ordered (perturbative) representations, the numerical solutions presented through Figures 1-3 do. 

$\eta (x,t)$ is the maximum height of the observed group wave, with a wavelength of $2\pi$ that gives the wave a steepness of $\pi$, for each harmonic, which is approximately 1/3 or 0.3 as predicted by Stokes and the steepness of the group wave formation will be closer to around 0.4. In order to examine the wave profile over a period, we also need to estimate the profiles for the acceleration and kinetic energy of the wave; starting from the expression in equation (\ref{n-order}), these will respectively be the first and second derivatives of the function $\eta(x,t)$ against the variable $\theta$, thereby leading to the following expressions for acceleration ($f$) and kinetic energy (KE) respectively as

\begin{subequations}
\begin{equation}
f=\frac{d \eta}{d \theta} = -a\sin(\theta) -\displaystyle \sum_{n=2}^{\infty} \zeta^{n} \left(\frac{n^2}{2^{n}}\right) \:\sin(n\theta),
\label{accn-eqn}
\end{equation}
\begin{equation}
\text{KE}=\frac{d^2 \eta}{d \theta^2} = -a\cos(\theta) -\displaystyle \sum_{n=2}^{\infty} \zeta^{n} \left(\frac{n^3}{2^{n}}\right) \:\cos(n\theta).
\label{ke-eqn}
\end{equation}
\end{subequations}

The above expressions depicted in equations (\ref{n-order}), (\ref{accn-eqn}) and (\ref{ke-eqn}) are the $n$-ordered sums respectively of the velocity, acceleration and kinetic energy of the Stokes surface wave. 

\begin{figure}[htbp]
        \includegraphics[width=0.49\textwidth]{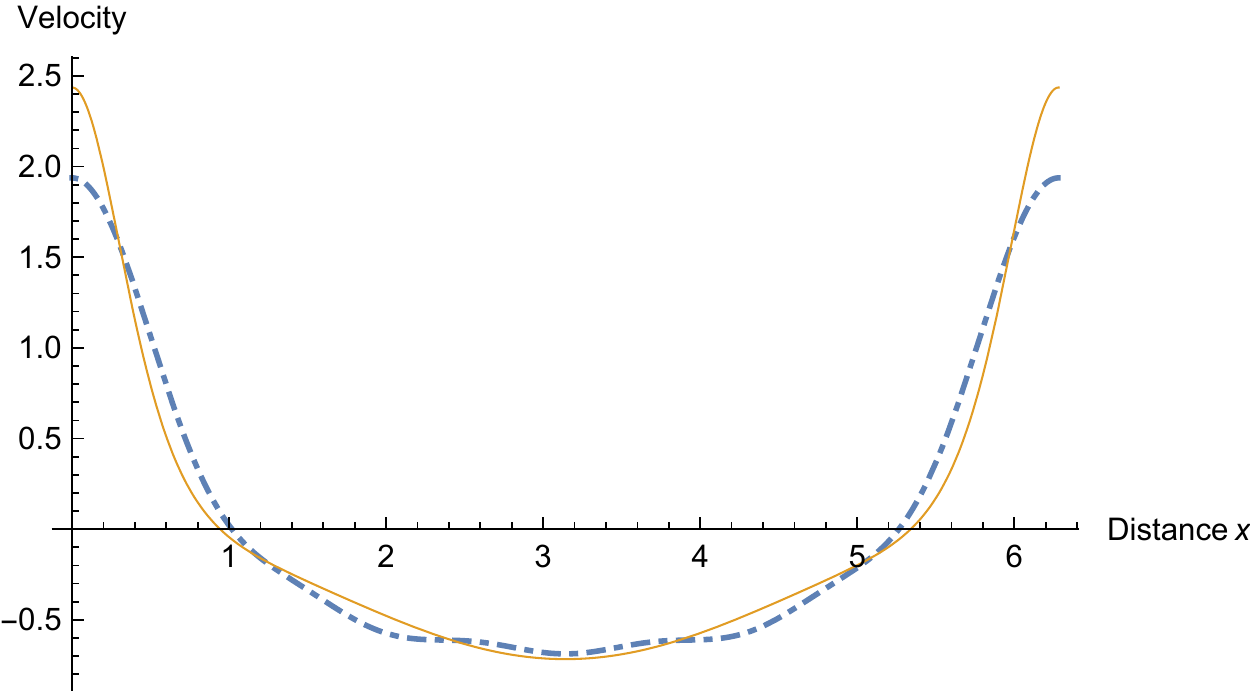}
        \caption{The velocity plotted against the spatial distance $x$ for $\zeta=0.99$. The solid line represents the exact n-order solution from equation (\ref{n-order}) while the dot-dashed line represents the approximate Fourier series solution up to fifth order of the Stokes approximation as given in equation (\ref{vel-stokes}).}
        \label{plotvel}
\end{figure}

Instead of this new representation, if we were to use the Fourier series representation due to Stokes, the corresponding forms for velocity, acceleration and kinetic energy will respectively have been

\begin{subequations}
\begin{equation}
v_{\text{Stokes}} = \displaystyle \sum_{n=1}^{n=5} 2^{n-1} \:\cos(n\theta) 
\label{vel-stokes}
\end{equation}
\begin{equation}
f_{\text{Stokes}} = -\displaystyle \sum_{n=1}^{n=5} n2^{n-1} \:\sin(n\theta) 
\label{accn-stokes}
\end{equation}
\begin{equation}
{\text{KE}}_{\text{Stokes}} = -\displaystyle \sum_{n=1}^{n=5} n^2\:2^{n-1} \:\cos(n\theta) 
\label{ke-stokes}
\end{equation}
\end{subequations}

\begin{figure}[htbp]
        \includegraphics[width=0.49\textwidth]{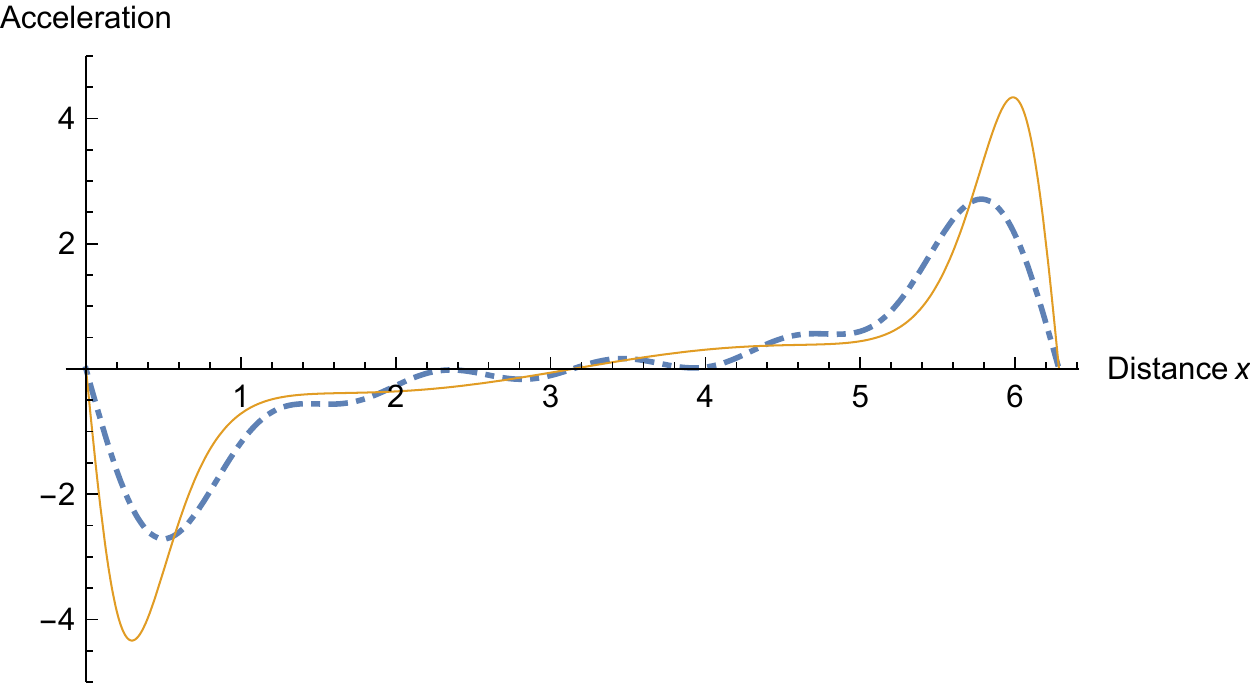}
        \caption{The acceleration plotted against the spatial distance $x$. The solid line represents the exact n-order solution from equation (\ref{accn-eqn}) while the dot-dashed line represents the approximate Fourier series solution up to fifth order of the Stokes approximation as given in equation (\ref{accn-stokes}) for $\zeta=0.99$.}
        \label{plotaccn}
\end{figure}

In the following, we compare the Stokes solutions shown in equation (\ref{vel-stokes}), (\ref{accn-stokes}), (\ref{ke-stokes}), going up to the fifth ordered expansions as in \cite{fenton1985} against the n-order accurate sum that we have propounded through equations (\ref{n-order}), (\ref{accn-eqn}) and (\ref{ke-eqn}) for $\zeta=0.99$.

\begin{figure}[htbp]
        \includegraphics[height=0.3\textheight,width=0.5\textwidth]{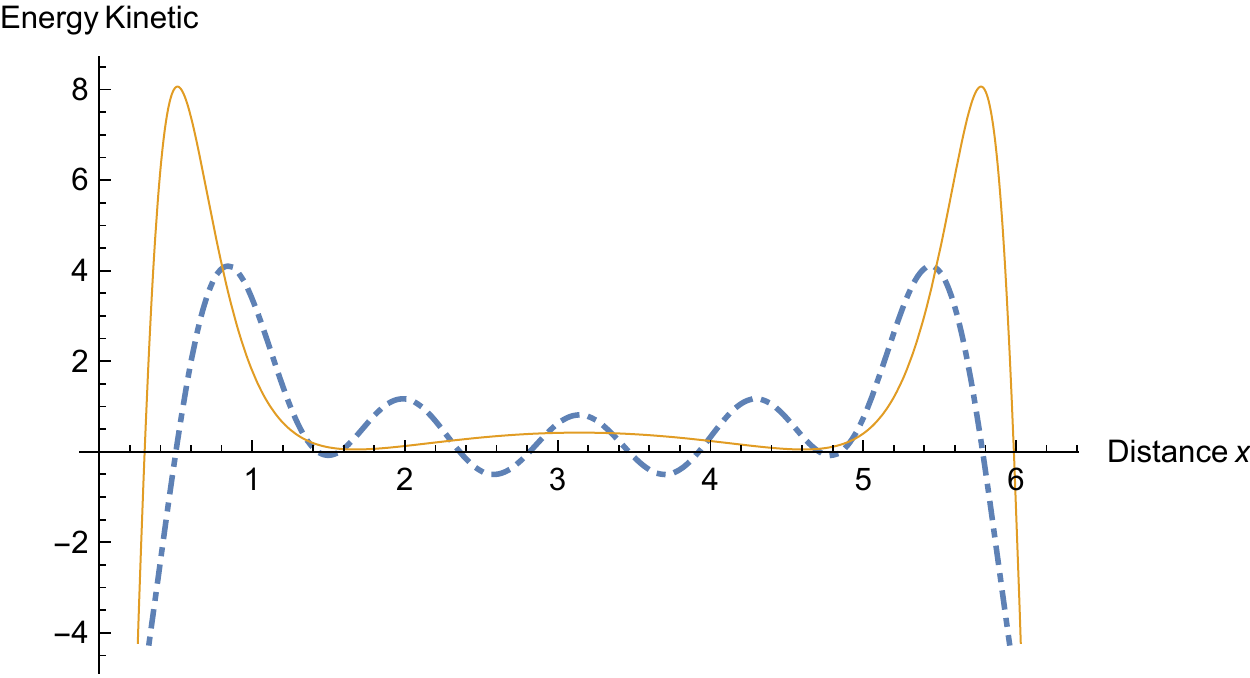}
        \caption{The kinetic energy plotted against the spatial distance $x$. The solid line represents the exact n-order solution from equation (\ref{ke-eqn}) while the dot-dashed line represents the approximate Fourier series solution up to fifth order of the Stokes approximation as given in equation (\ref{ke-stokes}) for $\zeta=0.99$.}
        \label{plotke}
\end{figure}

\newpage

As shown in Figures \ref{plotvel}, \ref{plotaccn} and \ref{plotke}, the solid lines respectively representing the velocity, acceleration and the kinetic energy from our new $n$-ordered solutions match closely with that of the equivalent quantities from Stokes Fourier series expansions, represented by the dot-dashed curves. The value for $\zeta=0.99$ chosen is indicative of the Fourier amplitude limit $\zeta \to 1$. It is easy to see that choosing a slightly different value of $\zeta$ away from this limit will result in minor aberrations from these almost perfect fits but will nevertheless not obfuscate the converging form.

To summarize, the $n$-ordered solution for the Stokes surface velocity wave as presented in equation (\ref{n-order}) and shown in closed form in equation (\ref{vel-inf}) provides a compact solution accurate to all orders compared to the approximate fifth order perturbative solution as given in \cite{fenton1985} or using Stokes original Fourier series piecewise continuity, as shown in equation (\ref{vel-stokes}). Such a closed form solution has the unique advantage of convergence for all values of $x$, thereby providing a generic hypothesized solution for all perturbative orders, that none of the previous perturbative or Fourier solutions could offer. This is much more than an aesthetic analytical insight in to a long standing non-convergent problem. As an example, we can now provide a closed form solution of the oscillating pressure gradient near a rigid boundary layer plate as a function of the closed form solution: $u_{\text{osc}}=u_o(x)[\cos(\omega t) - e^{-kz} \eta_{n\to \infty}]$, a much improved higher ordered accurate solution compared to the first order approximation as was previously presented in \cite{batchelor2000}. The  method presented here could also serve as a complementary approach to the more detailed, and hence tenuous, estimation of Stokes wave asymmetry, leading to chaoticity, in analyzing modulated Stokes flows in deep water \cite{ablowitz2000} or in instability prediction of geometries involving coaxially sheared cylinders \cite{lopez2014}.

\end{document}